\documentstyle[12pt,axodraw]{article}

\large
\newcommand{\be}{\begin{eqnarray}}
\newcommand{\ee}{\end{eqnarray}}

\newcommand\ie {{\it i.e. }}
\newcommand\eg {{\it e.g. }}

\newcommand\half{\frac 1 2 }
\newcommand\noi {\noindent}

\begin{document}
\setlength{\baselineskip}{21pt}
\pagestyle{empty}
\vfill
\eject
\begin{flushright} USITP-97-04 \\
SUNY-NTG-97-17 \\
April 1997
\end{flushright}
\vskip 1.5cm
\centerline{\Large\bf Real Time Correlators in Hot (2+1)d QCD}
\vskip 1.2 cm
\begin{center}
{\bf T.H. Hansson\footnote{Supported
by the Swedish Natural Science
Research Council.}, J. Wirstam }  \\
Institute of Theoretical Physics\\
University of Stockholm \\
Box 6730, S-113 85 Stockholm, Sweden \\
\vskip 3mm
{\bf I. Zahed\footnote{Supported
by the US Department of Energy under Grant  No. DE-FG02-88ER40388.}
 }\\
Department of Physics \\
SUNY at Stony Brook \\
Stony Brook, New York, 11794, USA \\
\end{center}
 
\vskip 1cm\noi
\centerline{\bf ABSTRACT}
\vskip .5cm
We use dimensional reduction techniques to relate real time finite $T$
correlation functions in (2+1) dimensional QCD to bound state
parameters in  a generalized 't Hooft model with an infinite number of
heavy quark and adjoint scalar fields. While static susceptibilities and correlation
functions of the DeTar type can be calculated using only the light (static) 
gluonic modes, the
dynamical correlators require the inclusion of the heavy modes. In particular we demonstrate
that the leading T perturbative result can be understood in terms of the bound states of
the  2d model and that consistency requires bound state trajectories composed of both
quarks and adjoint scalars. We also propose a non-perturbative expression for
the dynamical DeTar correlators at small spatial momenta.

\vfill
\eject
\pagestyle{plain}

\newcommand \qb {{\overline q}} 
\newcommand \etabar {\overline\eta}
\newcommand \psibar {\overline\psi}
\newcommand \thetaf [4] {\Theta\left[\begin{array}{c}#1 \\ #2
\end{array}\right] (#3 , #4) }
\newcommand \Dsl {D\!\!\!\!/}
\newcommand \dsl {{\partial}\!\!\!\!/}
\setlength{\baselineskip}{23pt}

\noindent{\bf 1. Introduction }
\renewcommand{\theequation}{1.\arabic{equation}}
\setcounter{equation}{0}
\vskip 5mm

About 10 years ago, DeTar \cite{detar} conjectured that the finite
T excitation spectrum of QCD was much richer than previously expected -
excitations with wavelengths longer than the magnetic scale $1/g^2 T$ are
color singlets, at the electric scale $1/gT$ there are colored plasmons and
at the hard scale $1/T$ colored quarks and gluons. He also
proposed to test this conjecture by studying color singlet current-current
correlators of the type
\be
G^{\alpha \alpha}(\vec x) = \frac{1}{\beta^2} \left\langle \int_{0}^{\beta}
d\tau \psibar(\vec x,\tau) \Gamma^{\alpha} \psi(\vec x,\tau) \hspace{0.2cm}
\int_{0}^{\beta}
d\tau '  \psibar(\vec 0,\tau ') \Gamma^{\alpha} \psi(\vec 0,\tau ')  \right\rangle \ \ \ ,
\ee
where $\langle \cdots \rangle$ is the average over the Gibbs ensemble. 
Lattice measurements of these ``DeTar correlators'' 
in Euclidean space, relevant for studying screening effects in the finite
temperature theory, suggest that the screening masses 
are about $2\pi T$ for the light vector and pseudovector excitations,
and smaller for the scalar and pseudoscalar. 

Some years ago, it was suggested \cite{two,KOCH} that the lattice results on the screening 
lengths and the spatial extension of the static correlators at high
temperature could be understood in terms of a dimensionally reduced theory which is 
known to confine in the high temperature limit. It was shown that the static
correlation functions can be calculated from an effective theory 
involving nonrelativistic quarks with colored Coulomb interactions. 
The screening lengths were found to asymptote the free quark values,
with spatially correlated quarks in the transverse directions at 
all temperatures. These ideas have been further developed in several
papers, notably those of Huang and Lissia \cite{huang1, huang2}, 
who have studied static properties at high temperature
in QCD and in the Gross-Neveu model.

In spite of our  increased understanding of the static aspects of high
temperature QCD, DeTar's main point, namely that the high
temperature phase of QCD is confined in the sense that all low energy
excitations are color singlets, is still neither proven nor
disproven. The main difficulty is that although there are many
sophisticated and efficient techniques for calculating {\em Euclidean}
correlation functions, there are no good methods, neither 
analytical nor numerical, to obtain the
{\em real time, finite T} correlators. In principle these can be obtained by
analytically continuing the Euclidean functions, as is done
routinely in perturbation theory, but on the lattice this is very
difficult without having a rather detailed {\em a priori} knowledge about the analytic
structure of the finite T spectral functions (for some alternative attempts 
in QCD see \cite{LARRY} and in simple model systems see \cite{realt1}). 

There are several lines of attack on this problem. The most direct is
perturbation theory, where there are well developed methods for 
calculating real time correlators \cite{pert}, but since QCD is known
to be nonperturbative at large distances even at infinite $T$, this method
is of limited value. Another way is to make some
specific assumption about the spectral functions, as for instance pole
dominance with $T$ dependent masses, and then use sum-rule techniques to check
the consistency of the assumptions and estimate spectral parameters in
terms of condensate values. This method is likely to be good at
temperatures well below $T_c$ \cite{bochsha1, bochsha2, dona, lee}, 
but has also been attempted at large $T$ \cite{sumrule2}.  
The last approach is to attempt exact, or
at least controlled, analytical calculations in simple models, with
the hope that some of the features can be generalized to realistic
cases. The present 
paper is devoted to a study of this kind using 2+1 dimensional QCD as a model
theory.  

The main idea of our calculation is to express  the real time finite $T$
correlators in 2+1 d QCD in terms of spectral parameters (bound state
masses, form factors) of a generalized 't Hooft model, which is obtained simply by
rewriting the theory in 2d language. This is of course only advantageous if the heavy
(\ie nonstatic) modes can be totally, or to a large extent, 
ignored in the high $T$ limit.
This is known to be the case for static observables, and we shall argue that for zero
two-momentum, $\vec q =\vec 0$, this is true also for certain dynamical correlation
functions in the sense that only the heavy modes of the quark fields have to be retained.
This claim is based on a comparison between the high temperature limit of perturbation
theory in the 2+1 d model and the generalized 't Hooft model. This comparison also
shows that for nonzero $\vec q$, the heavy scalars (originating from the non-static
gluon modes) must be retained, which implies that the two-dimensional method is of 
limited value. However, in this case we can show that consistency
requires the existence of 
linear trajectories of bound states containing heavy adjoint scalars in addition to
quarks and antiquarks. We shall briefly discuss this somewhat
surprising  result in relation
to the recent work on QCD$_2$ with adjoint scalars. 

In the next section we show how to reformulate 2+1 d QCD in a two dimensional language,
recall some basic results about the 't Hooft model and generalize certain formulas to
the case of unequal mass bound states. Section 3 is devoted to a study of the $\vec x$
and $t$ dependent quark susceptibility which is calculated both in perturbation theory
and from the 2d bound states. In   section 4 we suggest a
non-perturbative expression for non-static DeTar correlators at small
$\vec q$, and we close in section 5 with a discussion of our
results and some general comments.

\vskip 1cm
\noindent
{\bf 2. Dimensional reduction }
\renewcommand{\theequation}{2.\arabic{equation}}
\setcounter{equation}{0}
\vskip .5cm
\newcommand \xh {\hat x}
\newcommand \ah[1] {\hat A^{#1} } 
\newcommand \ph {\hat\psi}
\newcommand  \pbh {\hat{\bar\psi}}
\newcommand \ghu[1] {\hat\gamma^#1}
\newcommand \ghd[1] {\hat\gamma_#1}
\newcommand \om {\omega}
\newcommand \psib {\overline\psi}

Our starting point is the Euclidean action for massless QCD$_3$,
\be
S_{3E} = \int d^3\xh\, {\cal L}_{3E}(\ah \mu, \ph, \pbh) \ \ \ , \label{21}
\ee
where
\be
{\cal L}_{3E}(\ah \mu, \ph, \pbh) = \frac 12 {\rm Tr} \hat F_{\alpha\beta} \hat
   F_{\alpha\beta}  + \pbh(-i\ghd \alpha \hat\partial_\alpha + 
       g_3 \ghd \alpha {\hat A_{\alpha}})\ph \ \ \ ,  \label{22}
\ee
with the conventions $x^\alpha = (x^i,\tau)$, $\alpha, \beta
= 1,2,3$ and the gamma matrices are defined by
$\left( \ghu 1, \ghu 2, \ghu 3 \right) = \left(-\sigma^1,\sigma^2,
  \sigma^3 \right)$
with $\sigma^i$ the usual Pauli matrices. $g_3$ is the (dimensionful) coupling constant
in three dimensions.  
Next we simply rewrite the action in two-dimensional language, using
dimensional reduction techniques. We differ however from \eg
ref. \cite{two} in that we keep all the nonstatic modes as well. These heavy modes
will then manifest themselves as massive fields in the 2d action.
We start by making the following mode expansion  of the quark and gluon fields  
\be
\ph(\xh^i,\tau) &=& \sum_{n=-\infty}^{\infty} \ph_n(\xh^i) 
e^{-i\om_n^f\tau}   \nonumber \\ 
{\hat A^{\alpha}}(\xh^i,\tau) &=& \sum_{m=-\infty}^{\infty} {\hat A^{\alpha}_m}(\xh^i) 
e^{-i\om_m^b\tau} \nonumber \ \ \ , \label{24} 
\ee
where
$\om_n^f = \left( 2n+1 \right)\pi T$ and  
$\om_m^b = 2m\pi T$.
Using the static gauge condition $\partial_\tau \hat
A^3(\xh^i,\tau)=0$, which implies that the only remaining mode of the
$\hat A^3$ field is a (perturbatively) massless Higgs 
 $\hat A^3(\xh^i,\tau)= \hat A^3(\xh^i)= \phi(\xh^i)$, we get the
following final 2d form of the action, 
\be
S_{3E} &\rightarrow& S_{2E}^{YMH}[\hat A^j,\phi] - \beta\sum_n \int d^2\xh\ \pbh_n 
(\sigma^3\om_n^f + i\ghd j\hat\partial_j -g_3 \ghd j \hat A_j -
g_3 \sigma^3 \phi )\ph_n  + \nonumber \\ 
&+& \beta\sum_{\stackrel{\scriptstyle{n,m}}{n \neq m}}
\int d^2\xh\ \pbh_n g_3 \ghd j \hat A_j^{(n-m)} \ph_m   +
 S_{2E}^{NS}[\hat A_j,\hat A^n_j]  \ \ \ , \label{26}
\ee
where $j=1,2$ and $\beta = T^{-1}$.  $S_{2E}^{YMH}[A^i,\phi]$ is the action for a
two-dimensional Euclidean YM-Higgs 
model, as discussed in detail in \eg \cite{two, ymhiggs}, and $S_{2E}^{NS}[\hat A_j,\hat A^n_j]$
is 
the action for the nonstatic gluonic modes including the coupling to the static ones.
 
As discussed in \cite{two}, static DeTar correlation functions  can,
up to $O(1/T)$ 
corrections, be calculated from an effective dimensionally reduced
theory, although this is not possible in 3+1 d QCD. The only effect of
the heavy scalar fields is then to renormalize the coupling in the 2d model
and to add new vertices. One must of course keep the heavy quark
fields since there are no massless fermionic modes due to the 
anti-periodic boundary conditions. Even this model is
too difficult to treat analytically, and we will  make the simplifying
assumption that the scalar field $\phi$ can be neglected altogether. 
Arguments for why this does not change the basic properties of the
model were given in \cite{hans-rodd}, where it was argued that the net effect of
the scalar field is to renormalize the string tension and to introduce
an effective $1/R$ potential at large distances.
A strong indirect
evidence for the validity of this approximation was  given in
ref. \cite{zahed}, where it was shown that the
leading high temperature behavior of the susceptibility is correctly
reproduced by the contributions from 2d QCD bound states solely, neglecting any
effects of the Higgs field.

When it comes to nonstatic correlation functions the situation is much
more complicated. There is no {\em a priori} reason for neglecting the
heavy gluon modes, and as we shall see below in the case of the
non-static susceptibility, in general they do contribute.  

To apply the 2d QCD technology, we follow the approach in \cite{two} and
rotate to a fictitious 2d Minkowski space by: 
\be
\begin{array}{ccc}
  t\equiv x^0 = -i\xh^2 & \hat\partial_2 = -i\partial_0 & A_0 = i\hat A_2 \\
  \gamma^0 = \ghu 2 & \gamma^1 = i \ghu 1 & \{ \gamma^{\mu},
  \gamma^{\nu} \} = 2g^{\mu \nu}
\end{array} \label{27}
\ee
with $\mu, \nu =0,1$ and the Minkowski metric $(+ - )$.
We also perform a
chiral rotation on the quark fields and a rescaling (in order to get the mass term on the
conventional form and to give the quark fields the right canonical dimension \cite{two})
\be
\ph_n &=&\sqrt{T} e^{i\frac \pi 4 \sigma^3}   \psi_n \\ 
\pbh_n &=& i\sqrt{T} \psib_n e^{i\frac \pi 4 \sigma^3} .\nonumber \label{28}
\ee
With this, we get the following final form for the action we shall
use in the static sector,
\be
\tilde S_{3E} = -iS_{2M}^{YM}[A^\mu] - i \sum_n\int d^2 x\,  \psib_n 
          (i\gamma^{\mu} \hat\partial_{\mu} -g\gamma^{\mu}A_{\mu}
-M_{n} )\psi_n \ \ \ .  \label{final}
\ee
In (\ref{final}) the gluon field has been rescaled as well and $g=\sqrt{T}g_3$ is
the 2-dimensional coupling constant.
The model described by (\ref{final}) is QCD in two dimensions
with an infinite number 
of fermion fields with masses $M_n = (2n+1)\pi T$, which can be solved  in
the large $N_c$ limit, as originally shown 
by 't Hooft \cite{hooft} and  subsequently elaborated  by many others. We
will show that the methods developed in \eg \cite{callan} and \cite{enhorn} can be
adapted to the present problem, and that the difficulties associated with the
infinite number of quark fields can be handled.

The bilinear quark currents in the original 3d Euclidean theory takes
the following form, for the vector current $\hat V^{\alpha}$,
\be
\hat V^3 &=& -T\sum_{m,n}e^{-i(\omega_{n}- \omega_{m})
  \tau}\psibar_{m} \psi_{n} \label{vec3} \\
\hat V^j &=& \xi T \sum_{m,n}e^{-i(\omega_{n}- \omega_{m})
  \tau}\psibar_{m} \gamma^j \psi_{n} \label{currents}
\ee
with $j=1,2$ and $\xi = i^{(j-1)}$.

We now state some key results for the 't
Hooft model that we will use in the following. For a review and references
to original papers, see \eg \cite{ellis}. Due to the linear Coulomb
potential $V(r) = \sigma r$, with the string tension $\sigma = g^2N_c/\pi$,
there are no free quark states, but only color neutral hadrons, and at energies much larger
than the quark masses, the bound-state spectrum is approximately given by
\be
M_k^2 = \mu^{2}_{k}\sigma  = \pi^2 \sigma k \hspace{2cm} k \gg 1 \ \ \ . \label{spect}
\ee

In real QCD, the Greens functions for the vector
and axial vector currents are of special importance, since they couple to
external electro-magnetic and weak probes. In a model problem one is of course free to
consider any arbitrary Greens function, but we shall concentrate on the
vector current two point function, \ie the color singlet polarization tensor,
$\Pi_{\mu\nu}(\om,\vec q \,)$,  that at  $T=0$
determines the (would be) $e^+e^-$ production rate, at $T\neq 0$
and $\omega =0$ gives the screening of external electric and magnetic
fields, and for $\om\geq |\vec{q} \, |$ contains information about the finite $T$ quasi
particle spectrum. At $T=0$ this function has been extensively studied, and,
in the case of one quark flavor and to leading order in $1/N_c$, the
following exact expression has been derived \cite{hooft,callan,enhorn,ellis} 
\be 
\Pi_{\mu\nu}(\om,\vec q \,) = \left(q^2g_{\mu \nu} -
q_{\mu}q_{\nu}\right) \left( \frac{g^2 N_{c}}{\pi} \right)
\sum_{k}\frac{(g^{(V)}_{k})^2}{q^2- M_{k}^{2} +i\epsilon} \ \ \ ,     \label{pol}
\ee
where
\be
g^{(V)}_{n} = \int_{0}^{1} dx \, \phi_{n}(x) \ \ \ , \label{form}
\ee
and $\phi_n (x)$ is the meson wave function and $M_k^2$ its invariant mass.
Note that the mesons have zero-width to leading order in $1/N_c$.

A similar analysis for the scalar current two point
function in the leading $N_{c}$ approximation yields
\be
\Pi_S (\mu^2) = -\left( \frac{g^2 N_c}{\pi} \right) \sum_n 
\frac{(g_{n}^{(S)})^2}{\mu^2 -
  \mu^{2}_{n} +i\epsilon}  \ \ \ ,   \label{tpf1}  
\ee
where $\mu$, $ \mu_{n}$ is the four momentum of the current
and the bound state mass, in units of $g^2 N_c / \pi$,
with the coupling of the the bound state to the respective source given by
\be
g_{n}^{(S)} = \frac{1}{2}\int_{0}^{1} dx \, \left( \frac{\sqrt{\gamma_a}}{x} -
\frac{\sqrt{\gamma_b}}{1-x} \right) \phi_{n}(x) \label{formf} \ \ \ .
\ee
$\gamma_a, \, \gamma_b$ are the quark masses, also in units of $g^2 N_c / \pi$.
However, due to parity,
\be
m_a \int_{0}^{1} \frac{dx}{x} \, \phi_{n}(x) =
 \,  (-1)^{n+1} m_b \int_{0}^{1} \frac{dx}{1-x}  \, \phi_{n}(x) \ \ \ ,
\ee
so the sum over the bound states in eq. (\ref{tpf1}) is
effectively restricted to even $n$.
One can also show that, asymptotically,
\be
g_{n}^{(S)}  \rightarrow \frac{\pi}{\sqrt{2}} \ \ \ . \label{rform}
\ee
Similar relations can easily be derived for the pseudoscalar current.

For nonstatic correlation functions the situation is more
complicated. First, we must extend  the above treatment  to include bound
states of quarks with different masses. To do this we have generalized
the formulas (\ref{pol}) and (\ref{tpf1}) to the case of 
unequal masses, \ie to couplings of the form $\psib_n\gamma_\mu\psi_m$
where the field $\psi_n$ has mass $M_n$. The resulting expression for
the polarization tensor is identical to (\ref{pol}), but with the wave
functions being solutions to the unequal mass 't Hooft
equation; the same holds for the scalar correlator (\ref{tpf1}). 
Although the result is simple and intuitive, the derivation again
involves several nontrivial cancellations. There is however a more
subtle issue that regards the nonstatic gluon modes. As seen from
(\ref{26}), they manifest themselves in the 2d theory as heavy vector
fields in the adjoint representation. Since there is no physical light
gluon in two dimensions, we expect these heavy vector particles to be confined,
with a string tension $\sigma_{Ad} =  (C_{Ad}/C_F)\sigma$, where
$C_{Ad}$ and  $ C_F$ are the Casimir operators in the adjoint and
fundamental representations respectively. For a general discussion of 2d Yang-Mills
theory with adjoint matter fields, see \eg \cite{adj1, adj2,hans-rodd}. 
In the large $N_c$ limit we expect asymptotically straight trajectories
of stable bound states of the type
$\pbh_{n_1}A_{m_1}....A_{m_k}\ph_{n_2}$, and there is no reason to expect
that they should not contribute to correlation functions. In fact we
would expect a general formula of the form
\be
G(\omega, \vec q) = -\left( \frac{N_c}{\pi} \right) \sum_n \frac{1}{q^2 -
  (M^{B}_{n})^2} \left(g_{n}^{(B)}  \right)^2 \ \ \ ,   \label{vectors}  
\ee
where $M^{B}_{n}$ and $g_{n}^{(B)}$ are masses and couplings for these
trajectories. An important point, to be discussed below, is
that Lorentz invariance requires the vector particles to couple either
in pairs or via derivative couplings.

\vskip 1cm
\noindent
{\bf 3. Static and dynamical quark susceptibilities }
\renewcommand{\theequation}{3.\arabic{equation}}
\setcounter{equation}{0}
\vskip .5cm
In the large $N_c$ limit, all contributions to color singlet
correlation functions will be of the form given in (\ref{pol}) and (\ref{tpf1}). 
Using the results from the 't Hooft model, we can calculate 
the contribution from the $\qb q$ trajectory to the finite
temperature correlator of vector currents,
\be
G^{\mu\nu}(\Omega_N,\vec q \,) = \int_0^\beta d\tau\int d^2r\, e^{-i(\tau\Omega_N
+ \vec q\cdot \vec r)}\langle V^\mu(\tau,\vec r)V^\nu(0,\vec 0) \rangle \ \ \ , \label{gmn}
\ee
where the expectation value is with respect to the action $\tilde
S_{3E}$. We have set $\Omega_N= 2\pi N T$ and used $\vec r = (x,y)$. 
Using (\ref{final}) and (\ref{vec3}) we can relate $G^{33}$ to the 2d scalar correlator,
and by combining this with (\ref{tpf1}) we get
\be
G^{33}_{\qb q}(\Omega_N,\vec q \,)= -\left(\frac{g^2 N_c}{\pi} \right)^2 \sum_{n=-\infty}^\infty \sum_k
\frac {(g_k^{n+N,n})^2} {q^2 - M_k^2(n+N,n) + i\epsilon}\ \ \ , \label{gmnf}
\ee
where $M_k^2(m,n)$ is the mass and  $g_k^{m,n}$, given by (\ref{formf}),
the form factor of the k:th pole on the trajectory
describing mesons consisting of quarks with masses $(2n+1)\pi T$ and
$(2m+1)\pi T$. Similar expressions can be given for correlators of
other currents. Note that since we have rotated to a 2d Minkowski
space, $q^2 = q_0^2 - q_1^2$. After having evaluated (\ref{gmnf}) using 2d QCD
methods, we can obtain the corresponding real time finite $T$
correlation function by the analytic continuation $i\Omega_N = \omega +i\epsilon$. 

As shown in appendix A, perturbation theory to one loop gives the
following leading behavior at large T and $\left|\vec{q} \, \right| =q \gg \omega$,
\be
G^{33}(\omega,\vec q \,)&=& -\frac {g_{3}^{2} N_c} {\pi} \left(T\ln 2 
 - \frac {q\pi} {16}\right) + O(q^2/T)  \ \ \ \ \ \mbox{for} \ T\gg q 
 \gg \omega \ \ \ .\label{llim}
\ee
This expression is also
leading in $N_c$, so we can hope to reproduce it using the 't Hooft model, and
we shall now demonstrate that it can be
calculated from (\ref{gmnf}), \ie by
neglecting both the contributions from the bound states containing heavy
scalars, and the effect of the light Higgs field.   The derivation is a
straightforward extension of the calculation in \cite{zahed}, which we
now briefly recall. Since each term in the sum (\ref{gmnf}) is $\sim
O(1)$ in the temperature, a result $\sim T$ must emanate from the infinite range of the
summation. Thus we can use the asymptotic expressions (\ref{spect}) and
(\ref{rform}) for bound state masses and scalar form factors, and
furthermore replace the discrete sum with an integral of a continuous
bound state spectrum,
\be
\sum_k \rightarrow \frac 1 {\pi^2 \sigma} \int_{(m_1+m_2)^2} dM^2 \ \ \ ,
\ee
with $m_1$ and $m_2$ the two different quark masses.
The resulting expression for the relevant correlation function
 (after the rotation $q^2 \rightarrow - {\vec q}\,^2$) reads,
\be
G^{33}_{\qb q}(\omega,\vec q \,)= \frac {g_{3}^{2}N_c T} {2\pi} \sum_n\int_0^{{\Lambda}^2} \frac
{dM^2} {{\vec q}\,^2 + M^2 + (m_1 +m_2)^2} \ \ \ ,\label{ursp}
\ee
where we introduced an ultraviolet cutoff $\Lambda$. 
The static limit corresponds to $\Omega_N =  0$, where
$m_1^2 = m_2^2 = \omega_n^2$. Performing the $M^2$-integration and
rewriting the sum over $\omega_n$ as a countour integral we get for the $T$-dependent piece,
\be
G^{33}_{\qb q}(\omega,\vec q \,)=-\frac {g_{3}^{2}N_c}{ \pi} \int_{-i\infty +\epsilon}^{i\infty
+\epsilon}\frac {dz} {2\pi i}\, n_F(\beta z)\ln\left(\frac {\Lambda^2}
{q^2 - 4z^2}\right)\ \ \ ,
\ee
where $n_F$ is the Fermi distribution function. The logarithmic function has a cut along 
the real axis; in the positive real half-plane the cut starts at $q/2$. Following \cite{zahed} we
can deform and close the contour to get 
\be
G^{33}_{\qb q}(\omega=0,\vec q \,)= -\frac {g^{2}_{3}N_c}{\pi} \left( T\ln{2} 
-\frac q {4} \right) \label{qqstat}
\ee
and comparing this with (\ref{llim}) we see that the leading terms agree,
but the subleading ones deviate, although they have the same $k,T$ dependence
and the same sign.

We now turn to the non-static case. In appendix A we derive the
following perturbative result,
\be
G^{33}_{\qb q}(\omega,\vec q \,)&=& \frac {g^{2}_{3}N_c T} {2\pi} \ln 2 \left(\frac {q^2}
{\omega^2}\right) + O(Tq^4/{\omega}^4) \ \ \ \ \ \mbox{for}\ T\gg \omega \gg q
\label{flim} \ \ \ .
\ee
By comparing (\ref{flim}) with (\ref{llim}) we see that the two limits
$q=0 \ , \omega \rightarrow 0$ and $\omega = 0\ , q
\rightarrow 0$ do not commute, just as in 3+1 dimensions.

The limit (\ref{flim}) is a bit harder to reproduce with the 2-dimensional
technique than the static one, since $m_1 \neq m_2$ which
implies a more complicated analytic structure in the
$z$-plane. Performing the $M^2$-integration we get
\be
G^{33}_{\qb q}(\omega,\vec q \,)=-\frac {g_{3}^{2}N_c}{ 2\pi} \int_C \frac {dz} {4\pi i}\,
\tanh \left(\frac{\beta z}{2} \right)
\ln \left[ q^2 +\left( \sqrt{-z^2} + \sqrt{-(z-\omega)^2} \right)^2 \right] \ \ \ , \label{start}
\ee
where we analytically continued back to real time by $i\Omega_N =
\omega+i\epsilon$, and where the integration contour $C$ and the
square root cuts in the $z$-plane are shown in fig. 2(a) in appendix B. 
After some exercise in complex analysis, which is outlined in 
appendix B, we get 
\be
G^{33}(\omega,\vec q \,) &=& \frac {g_{3}^{2}N_c}{\pi} \int_{0}^\infty \frac{dx}{2\pi i} n_F(\beta x)
\left[ \ln \left( q^2 +\Omega_{N}^{2} -4 x^2 -4 ix\Omega_N \right) \right. \nonumber \\
&-& \left. \ln \left(q^2 +\Omega_{N}^{2} -4 x^2+4 ix\Omega_N \right) \right] \ \ \ . \label{cuts}
\ee
Taking $\vec q =\vec 0$ in (\ref{cuts}) the logarithms cancel for $\Omega_N=0$, so we conclude that in the
high-temperature limit,
\be
G^{33}_{\qb q}(\omega,\vec 0) \sim \omega + O(1/\omega^2) 
\ \ \ \  \mbox{for}\  \omega\ll T \ \ \ . \label{qqnonstat}
\ee
Note that the non-analytic properties of the perturbative results are also manifest
in (\ref{start}), since if we take the limits in reverse order we get the logarithmic cut
leading to (\ref{qqstat}), instead of  (\ref{qqnonstat}).
We have thus reproduced the nonleading behavior for the perturbative result (\ref{flim}),
in the limit $\vec q =\vec 0$, but from the expression (\ref{cuts}) it is
also clear that for $\vec q \neq \vec 0$
the leading term will not pick up any linear $T$-dependent piece, but only $O(q^2/ \omega)$.

At the end of the previous section, we pointed out that there is no
compelling reason to neglect the contribution from trajectories
containing heavy vector particles. We now want to argue that the
leading term in (\ref{flim}) is a likely sign of such a trajectory. The
simplest possible contribution would come from bound states of the
type $\pbh_{m_1}A_{n}\ph_{m_2}$ as illustrated in fig. 1. Since $A_n$
carries a vector index the coupling must be derivative, \ie  of the form $g_{\qb A q}
q^\mu$, where the coupling constant $g_{\qb A q}$ has dimension
1/mass. Without solving the resulting 3-body problem we can not
calculate this coupling, but we can nevertheless predict that the
contribution to the correlation function from the trajectory with
$m_1=m_2$ and $m_N = \Omega_N =2\pi N T$ will be 
\be
G^{33}_{\qb A q}(\omega,\vec q \,)= g_{\qb A q}^2 \frac{q^2}{(g^{(S)})^2}
G^{33}_{\qb q}(0,\vec q \,) = -g_{\qb A q}^2 q^2 
\frac {2 N_c T}{{\pi}^3} \ln 2 \ \ \ . \label{nonstatcorr} 
\ee
Although we have not determined the coupling, the presence of the
factor $T \ln 2$ in this formula is very suggestive when comparing to
the perturbative result (\ref{flim}). It is very natural to expect it to
have the same origin, \ie from an infinite summation over states on a
linear trajectory. For (\ref{nonstatcorr}) to be consistent with
(\ref{flim}) we must have $g_{\qb A q}\sim 1/m_n$, \ie the scale is set
by the mass of the scalar particle. We have not been able to prove
this.

\begin{center} \begin{picture}(300,80)(0,40)
\ArrowArc(150,50)(50,30,150)
\ArrowArc(150,100)(50,210,330)
\GCirc(108,75){3}{0.5}
\GCirc(192,75){3}{0.5}
\Gluon(108,75)(192,75){4}{8}
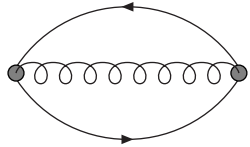
\end{picture} \\ Fig. 1. {\sl Contribution to the bound states containing a heavy vector 
particle.} \end{center}

To summarize this section, we have found a rather pleasing picture in terms of the
interplay (high temperature duality)
between perturbation theory and the dimensionally reduced theory
at high temperature. In the static
sector only the lightest gluon modes need to be retained in order to correctly reproduce
the screening length, whereas the non-static sector for consistency requires bound states
containing heavy vector particles.

\vskip 1cm
\noindent
{\bf 4. DeTar correlators at small $\vec q .$ }
\renewcommand{\theequation}{4.\arabic{equation}}
\setcounter{equation}{0}
\vskip .5cm
We argued above that the formula (\ref{gmnf}) can be used to
calculate the non-static DeTar correlator in the limit 
$\vec q =\vec 0$. However, to be able to make the final
analytical continuation $\Omega_N = -i\omega$ we need
expressions for both bound state masses and form factors that are
{\em analytic} in the masses, \ie in $\Omega_N$. Clearly some approximation
is needed, and we have tried to use the semiclassical methods
developed in \cite{wkb}. Even this turns out to be  too difficult,
since  the WKB expressions for the wave functions in general are very complicated, and
do not allow us  to derive explicit analytic expressions for the
form factors. However,  the nonrelativistic 
system of two particles with masses $m_1$ and
$m_2$ moving in a  linear potential $V(r) = \sigma r$ is relatively
easy to handle. The bound states of this system are given by,
\be
M_{k}^{2}(m_1,m_2) = M^2 + 2\lambda_k M \frac {\sigma^{2/3}} {(2m)^{1/3}} +
O(m_{1,2}^{-4/3}) \ \ \, 
\ee
for $\sigma \ll m_1\, , \, m_2$, with  $M=m_1 + m_2$, $m = m_1
m_2/M$ and
 $\lambda_n$ the n$^{th}$ root of the Airy function Ai$^\prime(x)$. 
We can also calculate the nonrelativistic form factors which are proportional to the
wave function at the origin, 
\be
\left(g_k^{n_1, n_2}\right)^2  \sim |\psi_k(0)|^2 \sim (m\sigma)^{1/3} \ \ \, \label{origin}
\ee
for even parity states. For odd parity states, the wave function
vanishes at the origin.  Although in
principle straightforward, we shall not elaborate  on the
evaluation of the normalization in (\ref{origin}). 
We are interested in the result after the analytic continuation 
$\Omega_N = -i\omega$, when the masses become,
\be
m_1 &=&  |2n+1| \pi T \\       \nonumber
m_2 &=& |i\omega +(2n+1)\pi T| \ \ \ ,\nonumber \label{masses}
\ee
where $|x|=(x^2)^\half$  which implies that ${\rm Re}(m_1+m_2) =
|2n+1| 2\pi T$. In particular,  we get for the lowest masses,
\be
M_k = 2\pi T \pm i\omega + \frac {\lambda_k \sigma^{2/3}} {(2\pi T)^{1/3}}
\left(1\pm \frac { i\pi T \omega} {(\pi T)^2 + \omega^2}\right) \ \ \ ,\label{mesonmass}
\ee
where the upper and lower sign refers to $n=0$ and $n=-1$
respectively. 

The question now is whether we can use
this information to calculate the correlator (\ref{gmn}), at least in
some specific kinematic region. This is not a simple problem. First
notice that the correlation function cannot be calculated directly for fixed
$\omega$ since the sum over the radial states does not converge fast
enough for the nonrelativistic approximation to be useful. Also, the
analytic structure in $\omega$ obtained from (\ref{cuts}) and
(\ref{nonstatcorr}) is too complicated to allow us to evaluate the 
real time correlator by a fourier transformation. The way to proceed
is instead to consider the following Fourier transform,
\be
\tilde G^{\mu\nu}(\Omega_N,x) = i\int \frac {dq_2} {2\pi}e^{iq_2 x}
G^{\mu\nu}(\Omega_N,0,q_2)
\ee
and calculate the contribution from the $\qb q$-trajectory only, to get
\be
\tilde G^{11}_{\qb q}(\Omega_N,x) = -i \frac {g^2N}{2 \pi}
\sum_{n=-\infty}^\infty \sum_k
(g_k^{n+N,n})^2 M_k(n+N,n)e^{-M_k(n+N,n)x} \ \ \ . \label{gmnf1}
\ee
In the limit $xT\gg 1$ and $g^2\ll T$, the expression (\ref{gmnf1}) is saturated
with a few nonrelativistic bound states. Note that in the static
limit $\omega =0$ we retain the static DeTar 
correlators derived in \cite{two}, and the main result of this section is
that the $x-$dependence of the dynamical correlators are obtained by a
straightforward extension of the earlier results. Also note that large
$x$ corresponds to small $q$. Following the arguments of the
previous sections, we expect trajectories containing heavy bosons to
give only small contributions.

We shall now make the
conditions that allow us  to saturate (\ref{gmnf1}) with only a few
bound states more precise. As already stated, $M_k$ depends linearly on $|n|T$,
so for $xT\gg 1$ equation (\ref{gmnf1}) will be dominated by the two lowest masses,
corresponding to (\ref{mesonmass}).
To suppress higher radial excitations we must require
\be
x\gg \left(\frac T {\sigma^2}\right)^{\frac 1 3} \ \ \ ,
\ee
which combined with the weak coupling condition $T\gg g \sim
\sqrt\sigma$, implies
\be
xT \gg \left(\frac {T^2} {\sigma}\right)^{\frac 2 3} >>> 1 \ \ \ ,
\ee
which means that we are in a region where the correlation function is
very small.

\vskip 1cm
\noindent
{\bf 5. Discussion and Conclusions}
\renewcommand{\theequation}{5.\arabic{equation}}
\setcounter{equation}{0}
\vskip .5cm
\noindent

We have shown how to relate real time finite $T$ correlation functions
in 3d QCD to quantities that can, at least in a certain limit, be
calculated analytically in a generalized 't Hooft model with an
infinite number of heavy quarks and scalars. 
Several assumptions and approximations
went into our calculation and we want to comment upon their validity. 

First, we neglected the Higgs field. This
is a questionable step, and might well be improved upon. The
effect of the Higgs field can be estimated in two limits. For
the stringlike excitations far up on the ``Regge'' trajectories we can
use the arguments of \cite{hans-rodd} to conclude that the only effect of
the Higgs field would be to renormalize the string tension and add a
non-leading $1/R$ term to the potential. This will clearly not change
the leading behavior of the spectrum which was the only thing needed
to retain the perturbative result for the susceptibility. For the
nonrelativistic bound states we can estimate the effect on the masses
from Higgs exchange to lowest order in perturbation theory, using an
effective four fermion potential of the kind discussed in \cite{three}. 
However, as long as there are no numerical simulations to compare with
it is hardly worth elaborating on corrections of this type. 

Secondly we want to emphasize a point already made in section 3. Starting from the
high temperature limit, we showed that consistency between the perturbative
high temperature theory (\ie 3d QCD) and the 
generalized 't Hooft model 
requires that the spectrum of the 
latter contains trajectories of bound states with heavy vector constituents, 
and that their decay strength decreases as $1/\omega$. These effects are 
dominant for time-like separations, whereas \eg the screening length can
be calculated by only taking into account the $\qb q$-trajectories.

The limited results of this paper may be compared with the very
detailed ones previously obtained in the Schwinger model \cite{schwinger} and the
large $N$ limit of the 2d Gross-Neveu model \cite{huang1}, where the analysis
is simpler. The Schwinger model is exactly solvable in terms of an elementary
scalar field, and the 2d Gross-Neveu model is tractable in the $1/N$ 
approximation. In contrast, 3d QCD studied here is rather similar to real QCD, 
and the difficulties are accordingly increased - in fact we find it 
encouraging that nonperturbative statements about the asymptotics of the Detar 
correlator, admittedly in some specific kinematical domain, can at all be made 
using analytical methods.

Finally we want to comment upon to which extent the results of this 3d model
can be generalized to real, (3+1)d QCD at finite temperature.
Starting from a high T phase, the correlators in 4d QCD show a similar non-analytical
behavior as encountered in 3d; in the static case we then have $G^{44}_{\qb q} \sim T^2$,
and in the nonstatic, $G^{44}_{\qb q} \sim T^2 (q^2/{\omega}^2)$. Since it is also
known that QCD in 3d is confining, it is from this point of view 
not unlikely that the dynamical correlators in the generalized 3d theory have to be
accompanied by a summation over
trajectories of bound states consisting of heavy vector fields as well, with a 
derivative coupling and a coupling constant $g_{\qb A q} \sim 1/ \omega$, just as 
in the 2d case. However, since 3d QCD is not solvable even in the
large $N$ limit,  this conjecture is not easy to verify. 
For the dynamical DeTar correlators, we expect the {\em asymptotic} $x$-behavior 
to be generic, {\em i.e.} that the sum over the bound states is saturated by only
a few nonrelativistic states and that the 
leading $\omega$ dependence of the correlation functions comes from
the  $\omega$ dependence of the  screening masses just like in (\ref{mesonmass})
and (\ref{gmnf1}). Again there will be corrections from the
mass-dependence of the binding energy. 

For {\em non-asymptotic} $x$, nothing can be said for sure, since many bound states will
contribute, and the nonrelativistic
approximation can no longer be trusted. One should note, however, that
to the extent that the constituent masses enter the total mass
additively, the leading $\omega$ dependence in  (\ref{gmnf1}) could still be
correct, since the factor $e^{i\omega x}$ is common to all the terms
in the sum. If valid, this argument, which is the same  in both three and four
dimensions, predicts a very simple  $\omega$ dependence for the DeTar
correlators, and might even be related to dynamical quark-gluon plasma
parameters. Again we stress that these comments on the non-asymptotic
behaviour are highly speculative.

\vskip 1.5cm
\noindent
{\bf Appendix A. The polarization tensor at high $T$ }
\renewcommand{\theequation}{A.\arabic{equation}}
\setcounter{equation}{0}
\vskip .5cm
\noindent
In this appendix we give the derivation of the asymptotic results, equations (\ref{llim})
and (\ref{flim}).
The polarization tensor $G(\Omega_N,\vec q \,) = (N_c g^{2}_{3})^{-1} G^{33}(\Omega_N,\vec q \,)$ is given by
\be
G(\Omega_N,\vec q \,) &=& T\sum_n \int \frac{d^2p}{{(2 \pi)}^2} {\rm Tr} \left({\sigma}_3 \frac{1}
{\not \hspace{-0.1cm}  p} {\sigma}_3 \frac{1}{\not \hspace{-0.1cm} p +\not \hspace{-0.1cm} q} \right)
\nonumber \\
&=& T\sum_n \int \frac{d^2p}{{(2 \pi)}^2}\left( \frac{p_{0}^2 +{\vec p}\,^2 +p_0 q_0 +\vec p \cdot \vec q}
{p^2 (p+q)^2} \right) \ \ \ ,
\ee
where $p_0 =(2 n +1) \pi T i$ and $q_0 =2 N \pi T i$.
Evaluating the frequency summation with the help of contour integration,
omitting the vacuum piece and performing the 
analytic continuation $i \Omega_N = \omega +i \epsilon$, we find for the imaginary part
\be
\hspace{-15.2cm} \lefteqn{{\rm Im} G(\omega,\vec q \,) = 
-\theta (q-\omega) \int_{q_+}^{\infty} \frac{dp}{2 \pi} \left[ \frac{4p^2
-4p\omega +\omega^2 -q^2}{\sqrt{4p^2 q^2 -(\omega^2 -q^2 -2p\omega)^2}} \right] \left( n_F (\beta (p-\omega)) - n_F (\beta p)
\right) }  \nonumber \\ \hspace{-15.2cm}
\lefteqn{- \theta (\omega-q) \int_{\omega_-}^{\omega_+} \frac{dp}{4 \pi} \left[ \frac{4p^2
-4p\omega +\omega^2 -q^2}{\sqrt{4p^2 q^2 -(\omega^2 -q^2 -2p\omega)^2}} \right] \left( n_F (\beta p) + n_F (\beta (\omega-p))
\right) \ \ \ ,} 
\ee
where $\theta$ is the step function. 
The integration limits are $q_+ = (q+\omega)/2$, $\omega_{\pm} = (\omega \pm q)/2$ 
and $q= \left| \vec q \, \right|$,
$p = \left| \vec p \, \right|$. From this expression it is clear that the imaginary part is proportional
to $\omega$ when $q \gg \omega$ and to $q$ for $\omega \gg q$.

The real part of the polarization tensor is
\be
-{\rm Re} G(\omega, \vec q \,) &=& \frac{T}{\pi} \ln 2 - \theta
(\omega-q)  
\int_{0}^{\infty} \frac{dp}{2 \pi} n_F (\beta p) \left( \frac{4p^2
+4p\omega 
+\omega^2 -q^2}{\sqrt{(\omega^2-q^2+2p\omega)^2 -4p^2 q^2}} \right)
\nonumber \\  
&&  + \theta (\omega-q) \int_{k_+}^{\infty} \frac{dp}{2 \pi} n_F
(\beta p) 
\left( \frac{4p^2 -4p\omega +\omega^2
-q^2}{\sqrt{(\omega^2-q^2-2p\omega)^2 
-4p^2 q^2}} \right) \nonumber \\  && -\theta (\omega-q)  
\int_{0}^{\omega_-} \frac{dp}{2 \pi} n_F (\beta p) \left( \frac{4p^2
-4p\omega 
+\omega^2 -q^2}{\sqrt{(\omega^2-q^2-2p\omega)^2 -4p^2 q^2}} \right)  
\nonumber \\ && + \theta (q-\omega) \int_{0}^{q_-} \frac{dp}{2 \pi} 
n_F (\beta p) \left( \frac{4p^2 +4p\omega 
+\omega^2 -q^2}{\sqrt{(\omega^2-q^2+2p\omega)^2 -4p^2 q^2}} \right)
\nonumber \\ && +
\theta (q-\omega) \int_{0}^{q_+} \frac{dp}{2 \pi} n_F (\beta p) 
\left( \frac{4p^2 -4p\omega +\omega^2
-q^2}{\sqrt{(\omega^2-q^2-2p\omega)^2 
-4p^2 q^2}} \right) \ \ \ . \label{polt}
\ee
Expanding in powers of $q/\omega$, we find 
\be
{\rm Re} G(\omega, \vec q \,) \simeq \frac{T}{2 \pi} \left(
\frac{q^2}{\omega^2} 
\right) \ln 2
\ee
in the limit $T \gg \omega \gg q$.

In the other limit,  $T \gg q \gg \omega$, the expansion of (\ref{polt}) gives
\be
{\rm Re} G(\omega, \vec q \,) = -\frac{T}{\pi} \ln 2 +\frac{q}{16} \ \ \ . \label{mel1}
\ee
Equation (\ref{mel1}) gives the screening  mass
\be
m_{el}^2 = -g^{2}_{3}G(\omega, \vec q=\vec 0) = g^{2}_{3} \frac{T}{\pi} \ln 2 \ \ \ , \label{mel2}
\ee
a result that also follows
from the pressure of the free quark-antiquark gas, with a nonvanishing chemical potential.
\vskip 1.5cm
\noindent
{\bf Appendix B. The polarization tensor from 2d, $\vec q = \vec 0$}
\renewcommand{\theequation}{B.\arabic{equation}}
\setcounter{equation}{0}
\vskip .5cm
\noindent
Here we show that the polarization tensor calculated
from the 2d theory has no leading $T$-behavior for ${\vec q}\,^2=0$.
Starting from the expression (\ref{ursp}) and performing the $M^2$-integration,
we can rewrite the Matsubara summation as
\be
G(\Omega_N, \vec q \,) = -\frac{1}{2\pi} \int_C \frac{dz}{4 \pi i} \tanh 
\left( \frac{\beta z}{2} \right) \ln \left[ {\vec q}\,^2 +\left( \sqrt{-z^2} +\sqrt{-(z-i\Omega_N)^2} \right)^2 
\right] \ \ \ , \label{circ}
\ee
where the contour $C$ is as shown in fig. 2(a).
We have two different cuts for the square roots, shown as wiggly lines
(for definiteness we have chosen $\Omega_N > 0$); the cut for $z=i\Omega_N$
is always at an even multiple of $i \pi T$ and hence it will not interfere
with the poles of $\tanh(\beta z/2)$.
These cuts correspond to having the cut for $\sqrt{f(z)}$ along the negative
real axis in the complex $f$-plane.
\vspace{0.5cm}

\begin{center} \begin{picture}(130,130)(25,-20)
\LongArrow(0,60)(120,60)
\LongArrow(60,0)(60,120)
\ZigZag(10,60)(110,60){1}{20}
\Text(100,100)[bl]{z}
\Line(96,96)(106,96)
\Line(96,96)(96,106)
\ZigZag(10,80)(110,80){1}{20}
\Vertex(60,60){1}
\Vertex(60,80){1}

\Vertex(60,105){2}
\Vertex(60,75){2}
\Vertex(60,85){2}
\Vertex(60,95){2}
\Vertex(60,65){2}
\Vertex(60,55){2}
\Vertex(60,45){2}
\Vertex(60,35){2}
\Vertex(60,25){2}
\Vertex(60,15){2}

\ArrowLine(65,10)(65,53)
\ArrowLine(55,53)(55,10)
\CArc(60,53)(5,0,180)

\ArrowLine(65,67)(65,73)
\ArrowLine(55,73)(55,67)
\CArc(60,73)(5,0,180)
\CArc(60,67)(5,180,0)

\ArrowLine(65,87)(65,110)
\ArrowLine(55,110)(55,87)
\CArc(60,87)(5,180,0)
\Text(60,-10)[]{(a)}
\end{picture}
\begin{picture}(130,130)(-25,-20)
\LongArrow(0,60)(120,60)
\LongArrow(60,0)(60,120)
\ZigZag(10,60)(110,60){1}{20}
\Text(100,100)[bl]{z}
\Line(96,96)(106,96)
\Line(96,96)(96,106)
\ZigZag(10,80)(110,80){1}{20}
\Vertex(60,60){1}
\Vertex(60,80){1}
\Text(60,-10)[]{(b)}

\Vertex(60,105){2}
\Vertex(60,75){2}
\Vertex(60,85){2}
\Vertex(60,95){2}
\Vertex(60,65){2}
\Vertex(60,55){2}
\Vertex(60,45){2}
\Vertex(60,35){2}
\Vertex(60,25){2}
\Vertex(60,15){2}

\Line(10,58)(110,58)
\Line(110,62)(10,62)
\ArrowLine(40,55)(30,55)
\ArrowLine(80,65)(90,65)

\Line(10,78)(110,78)
\Line(110,82)(10,82)
\ArrowLine(40,75)(30,75)
\ArrowLine(80,85)(90,85)
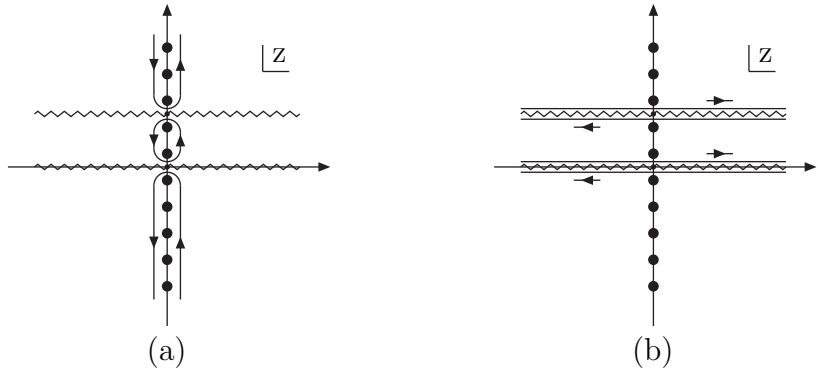
\end{picture}  \\ Fig. 2. {\sl The integration contour for $G^{33}$, before {\rm (a)} and 
after {\rm (b)} deformation.}
\end{center}

We can now deform the contour and arrive at
the contour shown in fig. 2(b), where we have neglected $T$-independent
pieces; this follows from the identity
\be
\frac{1}{2} \tanh \left( \frac{\beta z}{2} \right) = \pm \frac{1}{2} \mp
\frac{1}{e^{\pm \beta z} +1} \label{tanh} \ \ \ .
\ee 

The cut at $z=0$ then gives
\be
-\frac{1}{2 \pi} \int_{-\infty}^{\infty} \frac{dx}{4 \pi i} \tanh \left(
\frac{\beta x}{2} \right) \left[ \ln \left( {\vec q}\,^2 +\Omega_N^2 \right) -
\ln \left(-4x^2 +4ix\Omega_N +\Omega_N^2 +{\vec q}\,^2 \right) \right] \label{logg}
\ee
and by using (\ref{tanh}), equation (\ref{logg}) becomes for the temperature
dependent piece
\be
\frac{1}{2 \pi} \int_{0}^{\infty} \frac{dx}{2 \pi i}
 n_F (\beta x) \left[ \ln 
\left( -4x^2 -4ix\Omega_N +\Omega_N^2 +{\vec q}\,^2  \right) \right. \nonumber - \\ - \left.
\ln \left(-4x^2 +4ix\Omega_N +\Omega_N^2 +{\vec q}\,^2 \right) \right] \ \ \ .
\ee
In the same way, we find for the second cut at $z=i\Omega_N$,
\be
\frac{1}{2 \pi} \int_{0}^{\infty} \frac{dx}{2 \pi i}
n_F (\beta x) \left[ \ln 
\left( -4x^2 -4ix\Omega_N +\Omega_N^2 +{\vec q}\,^2 \right) \right. \nonumber - \\ - \left.
\ln \left(-4x^2 +4ix\Omega_N +\Omega_N^2 +{\vec q}\,^2 \right) \right]
\ee
by using
\be
\tanh \left( \frac{\beta x}{2} \right)
= \tanh \left( \frac{\beta (x +i \Omega_N )}{2} \right)
\ee
and so equation (\ref{circ}) becomes
\be
G(\Omega_N, \vec q \,) &=& \frac{1}{\pi} \int_{0}^{\infty} \frac{dx}{2 \pi i} n_F (\beta x) \left[ \ln 
\left( \Omega_N^2 +{\vec q}\,^2 -4x^2 -4ix\Omega_N  \right)  \right. \nonumber - \\ &-& \left.
\ln \left( \Omega_N^2 +{\vec q}\,^2 -4x^2 +4ix\Omega_N \right) \right] \ \ \ .
\ee
Taking $\vec q =\vec 0$
we see that by putting $\Omega_N =0$ the integral vanishes. Since it
also has to vanish for $\beta \rightarrow \infty$ we conclude that, after
analytic continuation,
\be
G^{33}(\omega, \vec q = \vec 0) \sim \omega \exp{\left(-\frac{\omega}{T}
\right)} \ \ \ ,
\ee
which is a subleading term and goes like $ \sim \omega$ for high temperatures.

\bibliographystyle{aip}
\bibliography{vacref}

\begin{thebibliography}{1}
 
\bibitem{detar}
C. DeTar,
\newblock Phys. Rev. {\bf D32} (1985) 276.

\bibitem{two}
T. H. Hansson and I. Zahed,
\newblock Nucl. Phys. {\bf B374} (1992) 177.

\bibitem{KOCH}
V. Koch, E. Shuryak, G. Brown and A. Jackson, 
Phys. Rev. {\bf D46} (1992) 3169; Erratum {\bf D47} (1993) 2157.


\bibitem{huang1}
S. Huang and M. Lissia,
\newblock Phys. Lett. {\bf B349} (1995) 484.

\bibitem{huang2}
S. Huang and M. Lissia,
\newblock Nucl. Phys. {\bf B480} (1996) 623.

\bibitem{LARRY}
D. Bodeker, L. McLerran and A. Smilga, Phys. Rev. {\bf D52} (1995) 4675;
A. Smilga, hep-ph/9612347

\bibitem{realt1}
S. Huang,
\newblock Phys. Rev. {\bf D47} (1993) 653.

\bibitem{pert}
J. I. Kapusta,
\newblock Finite-Temperature Field Theory, Cambridge University Press (1989);
A. L. Fetter and J. D. Walecka, 
\newblock Quantum Theory of Many-Particle Systems, McGraw-Hill (1971).

\bibitem{bochsha1}
A. I. Bochkarev and M. E. Shaposhnikov,
\newblock Phys. Lett. {\bf B145} (1984) 276.

\bibitem{bochsha2}
A. I. Bochkarev and M. E. Shaposhnikov,
\newblock Nucl. Phys. {\bf B268} (1986) 220.

\bibitem{dona}
H. G. Dosch and S. Narison,
\newblock Phys. Lett. {\bf B203} (1988) 155.

\bibitem{lee}
R. G. Furnstal, T. Hatsuda and S. H. Lee, Phys. Rev. {\bf D42} (1990) 1744;
C. Adami, T. Hatsuda and I. Zahed, Phys. Rev. {\bf D43} (1991) 921;
T. Hatsuda and S.H. Lee, Nucl. Phys. {\bf B394} (1993) 221;
M. Nowak, M. Rho and I. Zahed, `Chiral Nuclear Dynamics', World-Scientific
1997, and references therein. 

\bibitem{sumrule2}
T. H. Hansson and I. Zahed,
\newblock QCD Sum Rules at High Temperature, SUNY preprint (1990), unpublished;
I. Zahed,
\newblock Light relativistic bound states in hig temperature QCD, {\em in}
Thermal Field Theories, ed. H. Ezawa, T. Arimitsu, Y> Hashimoto
(North-Holland, Amsterdam 1991).
 
\bibitem{adj1}
S. Dalley and I. R. Klebanov,
\newblock Phys. Rev. {\bf D47} (1993) 2517

\bibitem{adj2}
G. Bhanot, K. Demeterfi and I. R. Klebanov,
\newblock Phys. Rev. {\bf D48} (1993) 4980; Nucl. Phys. {\bf B418} (1994) 15

\bibitem{ymhiggs}
E. D'Hoker,
\newblock Nucl. Phys. {\bf B138} (1978) 1.

\bibitem{hans-rodd}
T. H. Hansson and R. Tzani, 
\newblock Nucl. Phys. {\bf B435} (1995) 241.
 
\bibitem{zahed}
M. Prakash and I. Zahed,
\newblock Phys. Rev. Lett. {\bf 69} (1992) 3282.

\bibitem{hooft}
G. 't Hooft,
\newblock  Nucl. Phys. {\bf B72} (1974) 461; {\bf B75} (1974) 461.
 
\bibitem{callan}
C. G. Callan, N. Coote and D. J. Gross,
\newblock Phys. Rev. {\bf D13} (1976) 1649.
 
\bibitem{enhorn}
M. B. Einhorn,
\newblock Phys. Rev. {\bf D14} (1976) 3451.
  
\bibitem{ellis}
J. Ellis,
\newblock Acta Phys. Pol. {\bf B8} (1977) 1019.
 
\bibitem{wkb}
M. Prakash, M. Prakash and I. Zahed,
\newblock Ann. Phys. {\bf 221} (1993) 71.

\bibitem{three}
T. H. Hansson, M. Sporre and I. Zahed,
\newblock Nucl. Phys. {\bf B427} (1994) 545.
 
\bibitem{schwinger}
A. Fayyazuddin, T. H. Hansson, M. A. Nowak, J. J. M. Verbaarschot and
I. Zahed,
\newblock Nucl. Phys. {\bf B425} (1994) 553; J. V. Steele, J. J. M. 
Verbaarschot and I. Zahed,
\newblock Phys. Rev. {\bf D51} (1995) 1995.

\end{thebibliography}

\end{document}